\documentclass[aps,prb,twocolumn,superscriptaddress,floatfix,letter]{revtex4}
\usepackage{epsfig,amsmath,amssymb,color}
\begin{document}

\title{Universal emergence of the one-third plateau  in the magnetization process of frustrated quantum spin chains}
\author{F. Heidrich-Meisner} 
\author{I. A. Sergienko} 
\affiliation{Materials  Science and Technology Division, Oak Ridge National Laboratory,
 Oak Ridge, Tennessee, 37831, USA and\\
 Department of Physics and Astronomy, The University of Tennessee, Knoxville,
 Tennessee 37996, USA}
\author{A. E. Feiguin} 
\affiliation{Microsoft Project Q, The University of California at Santa Barbara, Santa Barbara, CA 93106, USA}
\author{E. R. Dagotto}
\affiliation{Materials  Science and Technology Division, Oak Ridge National Laboratory,
 Oak Ridge, Tennessee, 37831, USA and\\
 Department of Physics and Astronomy, The University of Tennessee, Knoxville,
 Tennessee 37996, USA}

\date{September 21, 2006}

\begin{abstract}
We present a numerical study of the magnetization process of frustrated quantum spin-$S$ chains 
with $S$=1, 3/2, 2 as well as  the classical limit. 
Using the exact diagonalization and density-matrix renormalization techniques, we 
provide evidence that a plateau at one third of the saturation magnetization 
exists in the magnetization curve of frustrated spin-$S$ chains with $S>1/2$. Similar to the case of $S$=1/2, this plateau
state breaks the translational symmetry of the Hamiltonian and realizes an {\it up-up-down} pattern in the 
spin component parallel to the external field. 
Our study further shows that this plateau exists both in the cases of an isotropic exchange and  in the easy-axis regime
for  spin-$S$=1, 3/2, and 2, but is absent in  classical frustrated spin chains with  isotropic interactions. 
We discuss the magnetic phase diagram of frustrated spin-1 and spin-3/2 chains as well as other emergent features
of the magnetization process such as  kink singularities, jumps, and even-odd effects.
A quantitative comparison of the one-third plateau in the easy-axis regime between spin-1 and spin-3/2 chains on the one hand
and the classical frustrated chain on the other hand indicates that the critical frustration and
the phase boundaries of this state rapidly approach the classical result as the spin $S$ increases.
\end{abstract}

\maketitle
\section{Introduction}
\label{sec:1}

The search for  novel quantum phases has been a strong motivation 
for theoretical investigations of quantum spin models.
Quantum states that do not  resemble classically ordered magnetic states may be realized
as a consequence of reduced dimensionality or competing interactions.
Famous examples are 
the Haldane phase of integer spin chains,\cite{haldane83}
or
the spin liquid ground states of spin-1/2  ladder systems,\cite{dagotto96} both of which exhibit 
a finite spin gap.\cite{dagotto92,white93a,golinelli94} 
The presence of competing interactions causes frustration and drives quantum phase transition,\cite{review-frust}
as for instance in the frustrated spin-1/2 chain that undergoes a transition from a critical, gapless phase into
a spontaneously dimerized one.\cite{okamoto92,white96}
Of great interest are the excitations above these ground states, which are reflected in
experimentally accessible quantities such as thermodynamic properties.

Here we are concerned with the interplay of frustration and an additional external parameter, the magnetic field. A magnetic field can close
zero-field gaps, but may as well induce gapped magnetic excitations in finite fields.\cite{oshikawa97,cabra97}
The magnetic phase diagram
can  experimentally be mapped out by studying the magnetization process, which theoretically
may be computed by means of  powerful and flexible numerical methods such as the density matrix renormalization group technique (DMRG).\cite{white92b,schollwoeck05}
One of the intriguing features of the magnetization curve of low-dimensional quantum magnets is the emergence
of plateaux at finite magnetic field, which indicate the presence of massive excitations. This phenomenon
has been predicted for several low-dimensional spin models.\cite{oshikawa97,cabra97,totsuka98,kuramoto98,nakano98,sakai98,sakai99,cabra00,
honecker00,okamoto01,
kitazawa00,sakai02,schulenburg02,okunishi03,okunishi03a,hida05,vekua06,laeuchli06,damle06}  
Experimentally, it has been observed in many quasi-low dimensional
magnetic materials,  realizing networks of spin-1/2 (Refs.~\onlinecite{nojiri88,kageyama99,miyahara99,kikuchi05, hase06}) 
as well as of spin-1 moments.\cite{narumi98}
It is the purpose of this work to study the magnetization process of frustrated quantum spin chains with  $S>1/2$
and establish the existence of a plateau state at one third of the full magnetization $M$.

The possible existence of magnetization plateaux in quantum spin chains was predicted
by Oshikawa, Yamanaka,  and  Affleck.\cite{oshikawa97} 
A notable result of this work provides a necessary condition
for the existence of plateau states at finite $M$:
\begin{equation}
p S (1-M)\in Z \,.
\label{eq:1}
\end{equation}
Here, $S$ is the spin, $p$ denotes the period (or length of the elementary unit cell) of the plateau state 
in real space. The theorem implies that $M$ always has a rational value on any plateau and it allows for translational 
invariance to be broken spontaneously, i.e., $p$ can be larger than 1. Note that 
unfrustrated spin-$S$ chains usually show a smooth magnetization curve, as has been known for quite some time 
already,\cite{parkinson85} implying that competing interactions beyond a simple nearest neighbor model,
  anisotropies, or special geometries
 are responsible for the plateau formation.
Indeed, theoretically, the existence of plateaux has been    established 
for, e.g.,  $S$=3/2 chains with onsite anisotropy\cite{oshikawa97, sakai98, kitazawa00} at $M$=1/2,
in frustrated and dimerized spin-1/2 chains at $M$=1/4,\cite{totsuka98}  three leg ladders at $M$=1/3,\cite{cabra97}
or exotic models such as the orthogonal dimer chain.\cite{schulenburg02}

The simplest model with competing interactions in one dimension is a spin-1/2 chain with nearest-neighbor (NN) 
and next-nearest-neighbor (NNN) exchange interactions, also called a zigzag ladder. 
 Only quite recently the existence of a $M$=1/3 plateau in the magnetization curve of 
 frustrated antiferromagnetic $S$=1/2 chains has been revealed by DMRG 
 calculations.\cite{okunishi03} The plateau is accompanied by broken translational symmetry with a 
period $p=3$ and an {\it up-up-down} ({\it uud}) structure in the spin component parallel to the external field. 
While several higher-$S$ chain models have been studied in the context of 
magnetization plateaux,\cite{oshikawa97, nakano98,sakai98,okamoto01, kitazawa00,kaburagi04} 
the case of frustrated chains with $S>1/2$ has not been addressed in sufficient detail. 
In particular, the existence of the $M$=1/3 plateau in these systems, which we report in this work, has not been explored.
Our main result, based on exact diagonalization (ED), DMRG, classical Monte-Carlo simulations (MC), and spin-wave theory,  
is that the $M$=1/3 plateau is  realized in  frustrated quantum spin chains with $S>1/2$ in the case of 
isotropic exchange and in the easy-axis regime as we show for the specific examples of $S$=1, 3/2, and 2.
As for spin-$1/2$,\cite{okunishi03} this plateau always has broken translational symmetry with period $p=3$ and
the {\it uud} structure.

A $M$=1/3 plateau with this N\'eel-type of order emerges in the Ising-limit of both quantum\cite{morita72,muraoka96,kaburagi04} and classical 
frustrated chains (see Ref.~\onlinecite{miyashita86} anf further references therein).
While the classical frustrated chain does not exhibit a finite magnetization plateau in the case of isotropic interactions as we show in
this work,
the $M$=1/3 plateau 
is stable against  quantum fluctuations.  
The parameter  region, however, where 
such plateau exists for the case of isotropic interactions shrinks as the spin $S$ grows. 
As we compute the full magnetization curves, we also obtain information on other 
 intriguing features such as 
kink singularities, macroscopic jumps, and even-odd effects in the magnetization curve,
and a very rich phase diagram  is indeed found.
We present a qualitative discussion of the magnetic phase diagram of spin-1 and spin-3/2 chains, which extends
the picture emerging from previous  studies for zero\cite{tonegawa95,kolezhuk96,kaburagi99,hikihara00,hikihara01,murashima05} 
and finite magnetic fields.\cite{kolezhuk05}

In contrast to the case of isotropic interactions, we find that the $M$=1/3  plateau 
region prevails in 
a substantial part 
of the magnetic phase diagram in the easy-axis regime. Its  boundaries 
rapidly approach the classical result in the easy-axis regime.
To this end,  we  study the magnetic phase diagram
of the classical frustrated chain as well by means of MC simulations and linear 
spin-wave theory, aiming at a comparison of emergent phases
with the quantum cases of $S$=1 and $S$=3/2.

In a very  recent experimental study, a $M$=1/3 plateau has been  observed
in the   frustrated diamond 
spin-1/2 material Cu$_3$(CO$_3$)$_2$(OH)$_2$  (Ref.~\onlinecite{kikuchi05}).
The existence of a 1/3 plateau has also been reported for the 
spin-1/2 trimer compound Cu$_3$(P$_2$O$_6$OH)$_2$ (Ref.~\onlinecite{hase06}).
Our results may be of particular relevance for  materials that 
have been suggested to realize frustrated spin-1 chains.
One example is CaV$_2$O$_5$ with competing next and next-nearest neighbor antiferromagnetic interactions.\cite{kikuchi01}
 A promising family of materials  realizing higher spin-$S$
 zigzag chains is NaR(WO$_4$)$_2$  where R represents In, Sc ($S$=0), Cr ($S$=1/2), and V ($S$=0 or $S$=1).\cite{masuda02}
The exchange constants in the latter case are estimated to be of the order of $180$K,\cite{masuda02} which may be small enough to access
a substantial part of the magnetization curve.  
Finally, the formation of  spin-1 zigzag chains has recently been reported for Tl$_2$Ru$_2$O$_7$, for which
a Haldane gap of $110$K is found.\cite{lee06} We hope that our results will stimulate the search for the one-third plateau
in potential frustrated spin chain materials.


The rest of the paper is organized as follows. In Sec.~\ref{sec:model}, we introduce the model
and briefly discuss the numerical techniques employed in this work: DMRG, Lanczos, and MC simulations. 
In Sec.~\ref{sec:2}, we present our numerical results for
the magnetization process of  frustrated spin  chains with $S$=1, 3/2, and $2$ and isotropic exchange interactions. 
We demonstrate the existence of the $M$=1/3 plateau for these quantum spins and numerically
determine the phase boundaries of the plateau region for spin 1 and 3/2 in the field  
vs frustration  plane.
We study the magnetization process of spin-1 chains in more detail, focusing on (i) 
kink singularities in the magnetization curve, (ii) low-field phases, and  (iii) and the combined effect of frustration and
onsite anisotropies. The latter is important as most materials that realize Haldane chains have a substantial
single-ion anisotropy.
In Sec.~\ref{sec:3}, we turn to the case of anisotropic frustrated chains in the easy-axis region.
The main purpose of this section is to demonstrate how the $M$=1/3 plateau observed in frustrated quantum spin-$S$ chains
is connected to the classical limit. To this end, we first analyze a classical frustrated spin chain by means
of MC simulations, which allows us to map out the ground-state phase diagram of the classical model in a finite
magnetic field. Analytical results for several phase boundaries are presented.
We then perform DMRG calculations for frustrated spin-$1$ and spin-$3/2$ chains at 
(i) a fixed exchange anisotropy, varying the frustration, and (ii) 
at a fixed frustration, varying
the exchange anisotropy, and determine the phase boundaries of the $M$=1/3 state for these two cases.
A summary and conclusions are presented in Sec.~\ref{sec:3}.
An Appendix is devoted to the details of a linear spin-wave calculation for the $M$=1/3 plateau state in the classical
limit.
%
\section{Model and methods}
\label{sec:model}
The Hamiltonian of a frustrated spin-$S$ chain with  onsite anisotropy $D$ in a magnetic field $h$ reads:
\begin{equation}
H= \sum_i  \lbrack J_1\, \mathbf S_i \cdot \mathbf{S}_{i+1} + J_2\, \mathbf{S}_i \cdot \mathbf{S}_{i+2} - h S^z_i  +D (S^z_i)^2 
\rbrack \, ,
\label{eq:2}
\end{equation}
where $J_1 > 0$ and $J_2 > 0$ are the antiferromagnetic NN and NNN exchange integrals, respectively. 
Here, $\mathbf{S}_i=(S_i^x,S_i^y,S_i^z)$ denotes a spin-$S$ operator acting on site $i$. 
We also include an exchange anisotropy $\Delta$, the same for the NN and NNN interactions. Hence, in our notation, with 
the usual definition of raising and lowering operators $S_i^{\pm}=S_i^x\pm i S_i^y$:
\begin{equation}
\mathbf{S}_i\cdot  \mathbf{S}_{j} \equiv \frac{1}{2} (S_i^+ S_{j}^- + S_i^- S_{j}^+) + \Delta\, S_i^z  S_{j}^z\,.
\label{eq:3}
\end{equation}
We restrict our analysis to  $\Delta\geq 1$, and we set $J_1=1$ and $D=0$ unless stated otherwise. The 
following normalization for the magnetization is used [this also applies to Eq.~(\ref{eq:1})]:
\begin{equation}
M=\frac{S_{\mathrm{total}}^z}{S\,N}; \quad S_{\mathrm{total}}^z= \sum_i S_i^z\, ,
\end{equation}
where $N$ is the total number of sites in the chain. Thus, $M$=1 at the saturation field $h_{\mathrm{sat}}$.

We  compute the ground-state energies $E_0(S_{\mathrm{total}}^z, h=0)$ in subspaces labeled by  $S_{\mathrm{total}}^z$ 
on chains with periodic boundary conditions (PBC) using the Lanczos algorithm. The ground-state energies of substantially 
larger chains with open boundary conditions (OBC) are calculated with DMRG.
Typically, we keep up to $m=400$ states in our DMRG calculations. 
Then, we include the Zeeman term and obtain the field-dependent ground-state energies
\begin{equation}
E_0(S_{\mathrm{total}}^z, h)= E_0(S_{\mathrm{total}}^z, h=0) -  h\, S_{\mathrm{total}}^z \,.
\label{eq:4}
\end{equation}
The magnetization curves are constructed by solving the equations 
$E_0(S_{\mathrm{total}}^z, h_{\mathrm{step}})=E_0(S_{\mathrm{total}}^z+s, h_{\mathrm{step}})$, which define
those magnetic fields at which the magnetization increases from 
$M=S_{\mathrm{total}}^z/(N\, S)$ to $M'=(S_{\mathrm{total}}^z + s)/(N\, S)$.
Steps larger than $s$=1 may occur.

In the classical limit, we treat $\mathbf S_i$ as three-dimensional unit vectors given by the polar angle $\Theta_i$ and 
azimuthal angle $\phi_i$. For each value of $h$, we find the ground state of $H$ using the ``simulated annealing'' classical MC 
 algorithm. Starting from a random configuration of angles $\lbrace \Theta_i,\phi_i\rbrace$ and an initial inverse temperature 
$\beta = \beta_\text{in}$, we proceed as follows. We perform sweeps through the chain, and for  each site $i=1, \ldots\, N$, two steps 
are performed:
(i) The energy $E=E(\lbrace\Theta_i,\phi_i \rbrace)$ of the current configuration is calculated from Eq.~(\ref{eq:2}).
(ii) New angles $\Theta'_i = \Theta_i + r\cdot w$ and $\phi' = \phi +  r\cdot w$ are tried, 
where $r\in \lbrack 0,1\rbrack $ is a random number and $w$ is the maximal change allowed. The new angles are accepted with probability
$p = \min \{\exp [\beta(E-E')],1\}$, where $E'=E(\lbrace\Theta_i',\phi_i' \rbrace)$ is the energy of the configuration with $\Theta_i$ and $\phi_i$ replaced by 
$\Theta'_i$ and $\phi'_i$. After each sweep through  the chain, $\beta$ is multiplied by the factor 
$(\beta_\text{fi}/\beta_\text{in})^{1/G}$, where $\beta_\text{fi}$ is the final inverse temperature and
$G$ is the number of sweeps. $w$ is adjusted such that the acceptance ratio is kept close to 50\% for efficiency. 
Typically, we use $\beta_\text{in} = 0.5$, $\beta_\text{fi} = 10$, and $G = 10^5$. This \emph{slow-cooling} technique 
prevents the simulation from getting stuck in 
 local minima in the energy landscape in most cases. Sometimes, close to phase boundaries, the system gets trapped 
in local minima, but repeating the algorithm a few times for the same set of parameters usually solves the problem.
After $\beta_\text{fi}$ has been reached, we fine-tune the ground-state energy by repeating  steps (i) and (ii) with 
$\beta = \infty$ for several sweeps until the energy change after a sweep becomes negligible.

\begin{figure}[t]
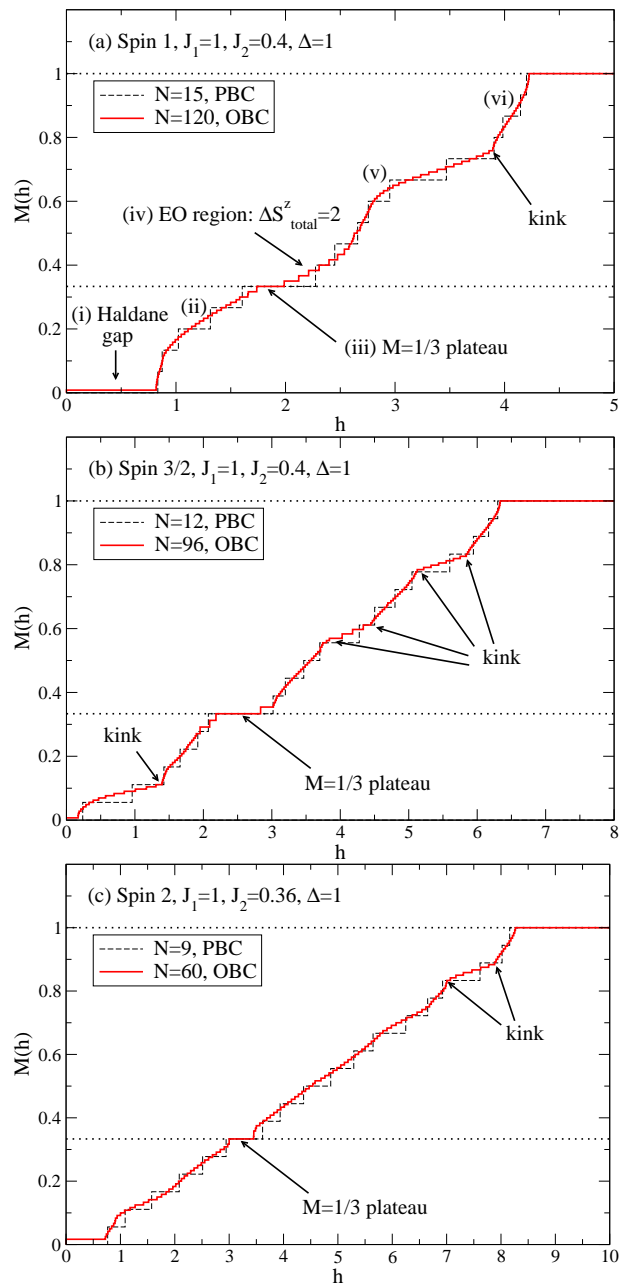

\centerline{\epsfig{file=figure1a.eps,angle=0,width=0.45\textwidth}}
\centerline{\epsfig{file=figure1b.eps,angle=0,width=0.45\textwidth}}
\centerline{\epsfig{file=figure1c.eps,angle=0,width=0.45\textwidth}}
\caption{(Color online) 
Magnetization curve $M(h)$ for  frustrated chains with an isotropic exchange ($\Delta=1$) 
 for (a) $S$=1 ($J_2=0.4$),   
(b) $S$=3/2 ($J_2=0.4$), and (c) $S$=2 ($J_2=0.36$). Solid(dashed) lines are  DMRG(ED) results. 
The horizontal dotted lines mark $M$=1/3 and $M$=1.}
\label{fig:1.1}
\end{figure}

\begin{figure}[ht]
\centerline{\epsfig{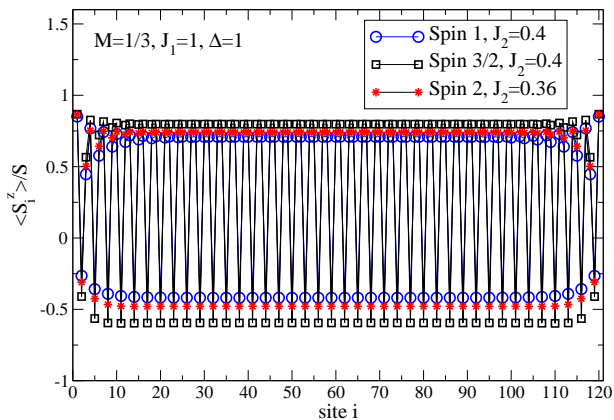}}
\caption{(Color online) 
 Onsite spin density $\langle S_i^z\rangle/S$ on the $M$=1/3 plateau state of frustrated spin-$S$ chains with $S$=1,3/2, and $2$
 and isotropic exchange interactions ($\Delta=1$) at $J_2=0.4$ ($S$=1 and 3/2) and $J_2=0.36$ ($S$=2).
 We show DMRG results for $N=120$ sites. 
 }
\label{fig:1.2}
\end{figure}

\section{Magnetization process of frustrated quantum spin chains with isotropic exchange interactions}
\label{sec:2}
In this section we present results for the magnetization curves of frustrated 
spin-$S$ chains with $\Delta=1$.
As a main result of this paper, we provide  numerical evidence that the 
 $M$=1/3 plateau, previously observed in the magnetization curve of frustrated spin-1/2 
chains,\cite{okunishi03,okunishi03a} exists for higher spins as well. The boundaries of the plateau phase strongly
depend on $S$, and the region in the $h$--$J_2$ plane where the $M$=1/3 state is realized
shrinks as a function of $S$. Our MC simulations in the classical limit show no evidence for a magnetization plateau in the
isotropic case.

\subsection{Magnetization curves of frustrated $S$=1, 3/2, and 2 chains} 
\label{sec:1.1}

The magnetization curves  at an intermediate frustration strength  are presented in Fig.~\ref{fig:1.1}
for the cases of spin 1 [Fig.~\ref{fig:1.1}(a), $J_2$=0.4], spin 3/2 [Fig.~\ref{fig:1.1}~(b),
$J_2$=0.4], and spin 2 [Fig.~\ref{fig:1.1}(c), $J_2$=0.36].
The magnetization plateau at $M$=1/3 is clearly observed in all three plots, and it can thus be considered a universal feature 
of frustrated quantum spin chains. 
In accordance with the criterion Eq.~(\ref{eq:1}) for the plateau formation, the translational invariance of the Hamiltonian is 
broken on the plateau, leading to a state with period $p=3$ and an {\it uud} pattern in the onsite spin density 
$\langle S_i^z\rangle$ parallel to the external field. 
The local spin densities corresponding to the plateau state are shown in Fig.~\ref{fig:1.2}.\cite{endnote63}
As we use OBC suited for DMRG calculations, one of the three possible patterns ({\it up-up-down, up-down-up, down-up-up}) is selected 
as a unique ground state due to energetically preferred orientations of edge spins, pointing up in our case.
We further observe deviations from the perfect {\it up-down-up} pattern, 
selected in our case, at the boundaries: edge-spin induced spin-spin correlations decay into the bulk.

Let us briefly comment on additional features of the magnetization curves presented in Fig.~\ref{fig:1.1}.
In the case of spin 1 shown in Fig.~\ref{fig:1.1}(a), we identify at least six regions, labeled (i)-(vi) in the viewgraph. First, the Haldane gap\cite{haldane83,white93a,kolezhuk96} shows 
up as a zero-field plateau.  A specialty of open spin-1 chains is the presence of 
effective spin-1/2 edge spins which have been studied in  detail both for unfrustrated\cite{white93a} and frustrated chains.\cite{kolezhuk96,roth98}
 These edge spins couple to the so-called Kennedy triplet, which in the thermodynamic limit is degenerate with 
 the ground state.\cite{kennedy90} The existence  of the effective free spin-1/2 edge spins emerges naturally within
 the AKLT (Affleck-Kennedy-Lieb-Tasaki) description of the ground state of a spin-1 chain.\cite{white93a,affleck87}
Within our numerical precision, this triplet causes  a zero-field magnetization of $M=1/N$ seen in 
the Haldane phase. Similar edge spin effects are present in all open spin-$S>1/2$ chains (see, e.g., Ref.~\onlinecite{roth98} 
and references therein): spin-3/2 chains have spin-1/2 edge spins whereas spin-2 chains have spin-1 edge spins, 
which consistently result in spurious $M=1/N$ and $M=2/N$ zero-field magnetizations 
seen in Figs.~\ref{fig:1.1}~(b) and \ref{fig:1.1}(c).

 At $h\approx 0.82$, the frustrated spin-1 chain enters a second region where the magnetization initially increases with a steep slope but
then follows $M(h)= \sqrt{h-b}$ with $b\approx 0.74$ for $h\gtrsim 0.9$.
The $M$=1/3 plateau is realized for $ 1.74 \lesssim h  \lesssim 1.99$. Note, however, that the phase boundaries of the plateau state cannot directly be obtained from $N=120$
sites for this frustration parameter, as finite-size effects are still significant in this regime. 
Both below and above the plateau (indicated as region iv), the magnetization increases in steps corresponding to $\Delta S^z_{\mathrm{total}}=2$. This may 
indicate binding effects 
between the elementary excitations in finite magnetic fields. 
In the high-field region below the saturation field, we highlight the 
presence of a {\it kink} in the magnetization curve at $h\approx 3.89$,
which correspond to jumps in the differential susceptibility $dM(h)/dh$. This kink splits the high-field region into parts 
(v) and (vi). A second, less obvious  kink may exist at smaller fields $h\approx 2.8$.
The kink singularities are related to  middle-field {\it cusp} singularities recently studied
in the literature.\cite{okunishi99,kolezhuk05,banks06}
We will discuss the kink singularities in more detail in Sec.~\ref{sec:phases_spin_1}.

\begin{table}[t]
\begin{ruledtabular}
\begin{tabular}{ccc}
Spin $S$ & $J_{2,\mathrm{crit,l}}$ & $J_{2,\mathrm{crit,u}}$ \\\hline
1/2 & 0.56 & 1.25 \\  
1   & 0.35 & 0.87\\   
3/2 & 0.34  & 0.78\\ 
2   & 0.35 & 0.72\\
$\infty$ & --  & --
\end{tabular}
\end{ruledtabular}
\caption{\label{tab:1.1} Isotropic exchange, $\Delta$=1: Lower and upper critical frustrations $J_{2,\mathrm{crit,l/u}}$ for the formation
of the $M$=1/3 plateau for spin 1/2 (from Ref.~\onlinecite{okunishi03}), 
spin 1, 3/2, and 2. }
\end{table}
 
Apart from the $M$=1/3 plateau, the prominent features of the magnetization curve of a 
frustrated spin-3/2 chain with $J_2=0.4$ are the zero-field gap
and several kinks. At zero magnetic field, the frustrated $S$=3/2 chain is gapless for $J_2 \lesssim 0.29$ and in a critical 
phase with antiferromagnetic correlations, while a gap opens beyond this value.\cite{roth98} 
This transition at zero field is of the Kosterlitz-Thouless type and is similar to the behavior of the frustrated $S$=1/2
chain.\cite{okamoto92,white96}
At larger $J_2\gtrsim 0.29$, the  system enters a dimerized phase, with a finite spin gap seen in Fig.~\ref{fig:1.1}~(b).
For the case of $h>0$,  several   kink singularities exist    in the magnetization process of spin-3/2 
chains both below and above the plateau, as is indicated in  Fig.~\ref{fig:1.1}(b). 
The derivative of the ground-state energies $E_0(S_{\mathrm{total}}^z,h=0)$ with respect to $M$
 exhibits discontinuous changes of the slope where the kinks emerge in the
magnetization curve, indicating at second-order transitions.
Similar to the case of spin 1, the magnetization increases in steps corresponding to $S_{\mathrm{total}}^z=3=2 S$
around the plateau.
 This hints at a very rich 
 phase diagram in a magnetic field that deserves to be explored in more detail.

The magnetization curve of frustrated spin-2 chains at $J_2=0.36$ exhibits
 a zero-field Haldane gap,\cite{hikihara01} similar to the case of spin 1, 
kinks at high fields, and the $M$=1/3 plateau.

\subsection{Magnetic phase diagram of frustrated spin-1 and 3/2 chains with isotropic exchange interactions: The $M$=1/3 plateau region}
\label{sec:phases_iso}

\begin{figure}[t]
\centerline{\epsfig{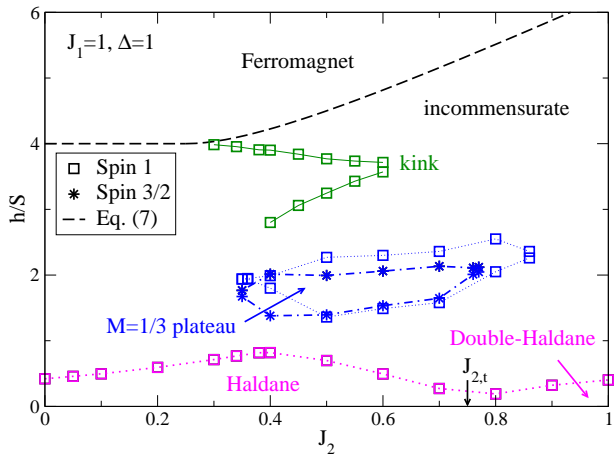}}
\caption{(Color online) 
Magnetic phase diagram of frustrated $S$=1 and 3/2 chains with an isotropic exchange ($\Delta=1$) as derived from ED and DMRG calculations 
of magnetization curves. Squares denote boundaries of $S$=1 phases, while stars are for $S$=3/2.
An arrow at $J_{2,\mathrm{t}}\approx 0.75$ at low fields   marks the first-order transition from the 
(gapped) Haldane phase found for $J_2 \lesssim 0.75$ to the likewise gapped double Haldane phase for the frustrated spin-1 chain.\cite{kolezhuk02}
}
\label{fig:1.4}
\end{figure}

 We next discuss how the boundaries of the plateau region depend on spin $S$. 
 First, we present results for the critical frustration for the plateau formation in Table~\ref{tab:1.1}.
 Second, the phase boundaries are determined as a function of frustration $J_2$ for spin 1 and spin 3/2
and these results are summarized in Fig.~\ref{fig:1.4}. Finally,
we present a  discussion of additional phases present in the case of spin $S$=1 in a separate section, Sec.~\ref{sec:phases_spin_1}.

Since the plateau is induced by frustration, we expect it to vanish as $J_2\to 0$, as well as
for $J_2\gg J_1$. In the latter case, the system separates into two decoupled unfrustrated NN chains.
For the extreme quantum case of $S$=1/2, DMRG studies\cite{okunishi03,okunishi03a}
report the existence of the plateau for $ 0.56\lesssim J_2/J_1 \lesssim 1.25$.
On the other hand, as mentioned earlier, our MC simulations for a classical frustrated spin chain
indicate the absence of the $M$=1/3 plateau in the isotropic case $\Delta=1$, thus we expect the 
region that allows for the {\it uud} state to become the ground state at $M$=1/3
to shrink with increasing spin $S$. 
We  therefore determine the critical lower and upper frustration $J_{2,\mathrm{crit,l/u}}(S)$ for the formation of
the plateau state. To this end, we  compute the onsite spin density $\langle S_i^z\rangle$ 
for several $J_2$, searching for the emergence of the {\it uud} pattern. We also
perform a finite-size scaling analysis of the plateau boundaries and compare chains with open and periodic
boundary conditions.
This procedure results in estimates of the critical frustrations for spin-1/2 consistent with Ref.~\onlinecite{okunishi03}.
Note that the precise determination of the critical frustrations is somtimes difficult as the gap, i.e., the width
of the plateau, is typically exponentially small close to the critical frustrations. Our results are intended to 
study trends and we estimate the critical frustrations listed in Table \ref{tab:1.1} and \ref{tab:2} to be correct with an error 
of $\pm 0.02$.

The results for $S$=1, 3/2, and 2 are summarized in Table~\ref{tab:1.1}, together
with data from Ref.~\onlinecite{okunishi03} for spin 1/2.
The data collected in Table~\ref{tab:1.1} renders $S$=1/2 a special case with critical frustrations  very different from $S>1/2$. 
For $S>1/2$, the critical lower 
$J_{2,\mathrm{crit,l}}(S)$ remains approximately constant, while $J_{2,\mathrm{crit,u}}(S)-J_{2,\mathrm{crit,l}}(S)$ shows 
a tendency to decrease with  increasing $S$, melting the plateau region from the large $J_2$ side.

\begin{figure}[t]
\centerline{\epsfig{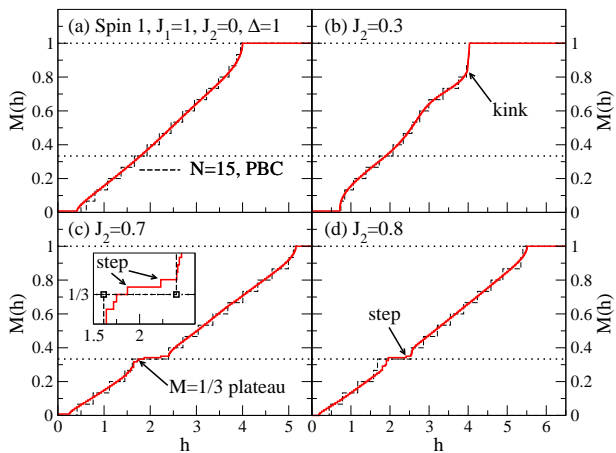}}
\caption{(Color online) 
Magnetization curves for  frustrated $S$=1 chains with an isotropic exchange ($\Delta=1$) for (a) $J_2=0$,
(b) $J_2=0.3$, (c) $J_2=0.7$, and (d) $J_2=0.8$. Full lines are DMRG results for $N=120$ sites, dashed lines
 are ED results for $N=15$. The inset in panel (c) is an enlarged view of the $M$=1/3 plateau. The squares
 indicate the boundaries of the plateau as seen on $N=15$ sites with PBC. }
\label{fig:1.3}
\end{figure}

\begin{figure}[t]
\centerline{\epsfig{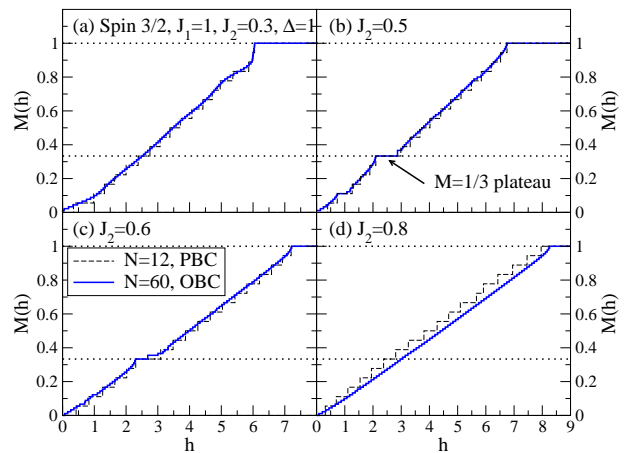}}
\caption{(Color online) 
Magnetization curves for  frustrated $S$=3/2 chains with an isotropic exchange ($\Delta=1$) for (a) $J_2=0.3$,
(b) $J_2=0.5$, (c) $J_2=0.6$, and (d) $J_2=0.8$. Full lines are DMRG results for $N=60$ sites, dashed 
lines are ED results for $N=12$  sites with PBC.}
\label{fig:1.3a}
\end{figure}

 We next elaborate more on the phase boundaries $h=h(J_2)$ of the plateau state in the 
$h$-$J_2$ plane for $S$=1 and 3/2. Our results from   DMRG calculations for several $J_2$ are summarized in Fig.~\ref{fig:1.4}, 
which shows the magnetic phase diagram of frustrated $S$=1 and 3/2 chains at $\Delta$=1.
Additional magnetization curves for $S$=1 and $J_2$=0.0, 0.3, 0.7, and 0.8 are shown in Fig.~\ref{fig:1.3} and 
in Fig.~\ref{fig:1.3a} for $S$=3/2 and $J_2$=0.3, 0.5, 0.6, and 0.8.

\begin{figure}[t]
\centerline{\epsfig{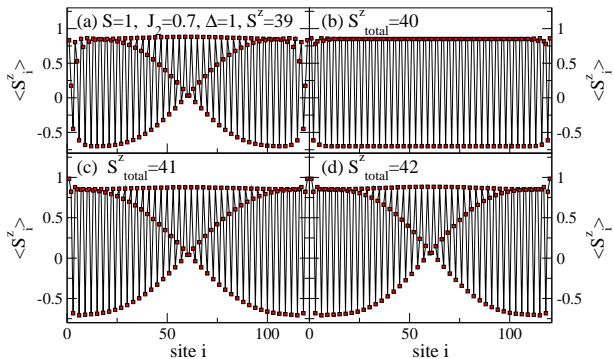}}
\caption{(Color online) 
Onsite spin density $\langle S_i^z\rangle$  state for $S$=1, $J_2=0.7$, $\Delta=1$, and $N=120$ 
sites (DMRG). 
Panels (a)--(d): $S^z_{\mathrm{total}}$=39--42. 
$S^z_{\mathrm{total}}=40$ corresponds to $M$=1/3.
}
\label{fig:1.5a}
\end{figure}

Close to the upper critical fields $J_{2,\mathrm{crit,u}}(S)$, the determination
of the boundaries is hampered by the existence of additional steps on the plateau. 
An example is shown in the inset of Fig.~\ref{fig:1.3}(c) for $J_2=0.7$, the steps are indicated by the arrow in the plot. Such steps have also been 
seen in the case of frustrated $S$=1/2 chains,\cite{okunishi03} and can be traced back
to the open boundary conditions. 
First, as we have verified for several cases, in which such steps are observed,
we find that 
 the step height scales down to zero with increasing system size. 
Consistent with the interpretation that the steps are boundary effects, 
they  are not seen if periodic boundary conditions are imposed.
Finally, we compare the onsite spin density 
$\langle S_i^z\rangle$ on the plateau to that of the step states, i.e., the magnetizations just below and the two magnetizations  just 
above the plateau. The spin densities are shown in Fig.~\ref{fig:1.5a} 
for $J_2=0.7$ and $N=120$ sites. 
On the plateau ($S^z_{\mathrm{total}}=40$), we find the {\it up-down-up}
pattern. For $S^z_{\mathrm{total}}=39,41,42$, i.e., just above and below
 the plateau, one-domain wall excitation sits in the middle of the
chain.\cite{okunishi03a} 

In the magnetic phase diagram Fig.~\ref{fig:1.4}, we scale the field with the value of $S$. 
We see that while the upper and lower  critical frustration $J_{2,\mathrm(l,u)}$ are similar for $S$=1 and 3/2,
the width and the critical field, i.e., the boundaries of the plateau region exhibit a significant
dependence on spin $S$.
In particular, we   emphasize the non-monotonic dependence of the plateau's width  
 on $S$ -- it is quite narrow for $S$=1/2,\cite{okunishi03} broader for $S$=1 and 3/2, and it disappears in the classical case 
$S=\infty$.


\subsection{Magnetic phase diagram of the frustrated spin-1 chain: Additional phases}
\label{sec:phases_spin_1}

Finally,
we present a  discussion of additional phases present in the case of spin $S$=1.
To our knowledge only the magnetization curve of the NN spin-1 chain has been computed (see, e.g., Ref.~\onlinecite{sakai91}).
Based on a field-theoretical analysis, a first suggestion for the magnetic 
phase diagram of frustrated $S$=1 chains has been put forward in Ref.~\onlinecite{kolezhuk05} for the low- and high-field
parts.
Our results basically confirm their picture and provide additional information on the so far unexplored middle-field region
where the $M$=1/3 plateau is found. 

Several studies of frustrated spin chains have focused on the zero-field 
phases,\cite{kolezhuk96,roth98,kolezhuk00,hikihara00,hikihara01,kolezhuk02,murashima05,kolezhuk05} with 
special interest in the emergence of chirality. 
As is well known, unfrustrated $S$=1 chains exhibit a Haldane gap at zero magnetic field, in contrast to half-integer spin 
chains that have gapless excitations.\cite{haldane83,white92b,golinelli94} As earlier studies have 
shown, frustrated $S$=1 chains have a gap for any $J_2$.\cite{rao94,allen95,kolezhuk96} 
The Haldane phase persists up to $J_2 \approx 0.75$, and is then followed by another gapped phase, the double-Haldane 
region.\cite{kolezhuk96} This transition is a consequence of the frustration driving the system from the  unfrustrated one-chain
($J_2\to 0$) to the two-chain limit ($J_2$=1).

The existence of a spin gap gives rise to a zero-field plateau in the magnetization curve, 
which is evident
from our data shown for $S$=1 and $J_2$=0.0, 0.3, 0.4, 0.7, 0.8 [Figs.~\ref{fig:1.3}(a), \ref{fig:1.3}(b), \ref{fig:1.1}(a), 
\ref{fig:1.3}(c), \ref{fig:1.3}(d)]. 
The zero-field magnetization of $M=S_{\mathrm{total}}^z=1/N$  due to the Kennedy triplet 
is seen in the Haldane phase,
but  disappears at the transition to the Double-Haldane phase.\cite{kolezhuk96}

We therefore estimate the spin gap  from the field where the total $S_{\mathrm{total}}^z$ increases  from 
$S_{\mathrm{total}}^z=1$(0) to $S_{\mathrm{total}}^z=2$(1) in the Haldane(Double-Haldane) phase.
Our  results for the spin gap are included 
in the magnetic phase diagram in Fig.~\ref{fig:1.4} (squares) and are in agreement  with the data of
Ref.~\onlinecite{kolezhuk02}.

While the magnetization curve of both the unfrustrated chain depicted in Fig.~\ref{fig:1.3}(a) 
and of chains with large $J_2$ (not shown in the figures) are mostly featureless, a more interesting behavior
arises in the range of parameters close to the plateau region.
Around the plateau itself we find evidence for a narrow region where the magnetization increases in steps corresponding to
$\Delta S^z_{\mathrm{total}}=2$: states with  an odd $S^z_{\mathrm{total}}$ never become ground states.
Such a region exists in frustrated $S$=1/2 chains as well, both in its anti- and ferromagnetic versions.\cite{okunishi03,hm06a} 
The corresponding state has been called the \emph{even-odd} (EO) phase.\cite{okunishi03} 
While the EO phase for $S$=1/2 is predominantly realized at large $J_2$, we observe the EO effects only in a narrow 
region around $J_2\approx 0.4$ [see Fig.\ref{fig:1.1}(a)]. 
  Starting from the limit of two 
decoupled chains, the existence of the EO regions can be motivated in a simple picture. When $J_1=0$, spins 
always flip in pairs, one on each chain.  The nontrivial question is whether a finite $J_1$ preserves this effect leading to 
pair-binding of excitations. This is indeed the case for the 
frustrated $S$=1/2 chain.\cite{okunishi03,hm06a} 
For spin 1,  however,  Kolezhuk and Vekua\cite{kolezhuk05}  conclude that an  EO phase  
is not possible in  frustrated chains.
Our DMRG results do not contradict this prediction as we find EO effects in a region that has not been studied in
Ref.~\onlinecite{kolezhuk05}.

To gain a qualitative understanding of the behavior in high magnetic fields, it is 
instructive to consider the 
dispersion of a single magnon, i.e, one spin flipped with respect to the fully polarized state.
The  dispersion $\epsilon_k$, generalizing results given in Refs.~\onlinecite{okunishi99} and \onlinecite{gerhardt98}, reads
\begin{equation}
\epsilon_k =  2 S \lbrack J_1 \cos k + J_2 \cos 2k - \Delta (J_1+J_2) \rbrack - h\,.
\label{eq:1.2}
\end{equation}
Here, and throughout, $k$ denotes the momentum. 
We also have included the exchange anisotropy $\Delta$ [see  Eq.~(\ref{eq:3})] for the sake of generality.
Equation \ref{eq:1.2} first allows us to derive the saturation field, which depends on $S$ 
only by an overall prefactor:\cite{gerhardt98}
\begin{eqnarray}
h_{\mathrm{sat}} &=&  2 S \lbrack J_1-J_2+\Delta(J_1+J_2) \rbrack \text{ for } J_2\le \frac{J_1}4,\nonumber \\
h_{\mathrm{sat}} &=& 2 S \left\lbrack \frac{J_1^2}{8 J_2}+J_2+\Delta(J_1+J_2)\right\rbrack \text{ for } J_2>\frac{J_1}4.
\label{eq:hsat}
\end{eqnarray}
The saturation field  is plotted as a dashed line in Fig.\ref{fig:1.4}. For $\Delta=1$ the saturation field is
independent of $J_2$ for $J_2\leq 1/4$.

For $J_2\gtrsim 0.25$,  a kink singularity splits  off the saturation field and 
divides the high-field region into two phases. A second kink emerges at lower fields
at $J_2\approx 0.4$ and then approaches the first one.
We observe the first high-field kink for $0.25\lesssim J_2\lesssim 0.6$, where it then 
merges with the second kink and eventually disappears, at least on the system sizes
studied here.
Just below the saturation field, where the one-magnon state is expected to give a correct description, 
we can understand its emergence along the lines of Ref.~\onlinecite{okunishi99}, where a similar effect has been observed 
for spin-1/2 chains.
It turns out that for $J_2 > 0.25$ the global minimum of $\epsilon_k$ at $k=\pi$ splits into 
 two local minima at the position of the high-field kink singularity. We can further conclude that 
 the high-field region between this kink and the saturation field is incommensurate with  $\cos k = -J_1/4 J_2$. 
 This agrees with Ref.~\onlinecite{kolezhuk05} where this part of the phase diagram has been suggested to 
 exhibit chiral order.
 
 We emphasize that the interpretation of the kink singularity  based on the one-magnon dispersion Eq.~(\ref{eq:1.2}) is justified 
 only just below the saturation field.  Indeed, even for the case of spin 1/2,  the 
 nature of the two phases left and right of the kink singularity 
  is currently under controversial discussion.
 According to Refs.~\onlinecite{okunishi03} and \onlinecite{okunishi99}, it signals a transition from a one-component 
Tomonaga-Luttinger liquid (TL1) to a two-component Tomonaga-Luttinger liquid (TL2) state. 
This interpretation  has recently been challenged
    by Kolezhuk and Vekua.\cite{kolezhuk05} Using bosonization techniques, they conclude that the  kink can be explained
    in terms of a renormalization of the magnetization due to an irrelevant operator
     in the incommensurate  region neighboring the TL2 (or chiral phase) 
    on the large $J_2$ side.

 Further studies are necessary to relate the three features of the $M(h)$ curves -- 
 the plateau, kink singularities, and 
EO effects -- with each other and should involve a detailed study of excitations of $S$=1 and 3/2 chains as well. 
Also, a numerical investigation of the high-field phases including the possible existence of  chiral order is a necessary step towards a 
complete understanding of the phases $S$=1 chains in finite magnetic fields.
 

\begin{figure}
\centerline{\epsfig{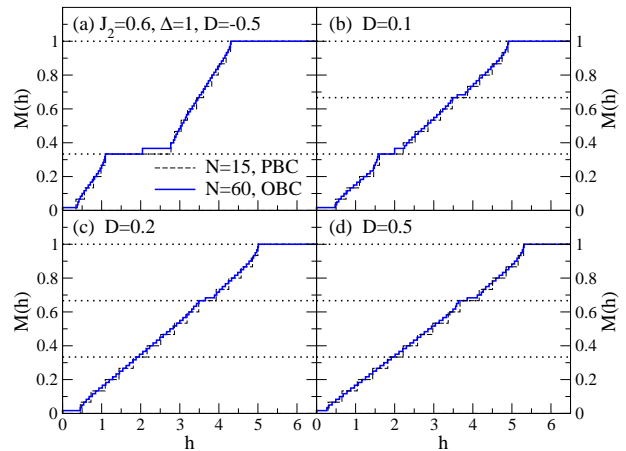}}
\caption{(Color online) 
Magnetization curves for the frustrated $S$=1 chain with isotropic exchange, $\Delta=1$ and $J_2=0.6$, and a finite onsite anisotropy.
(a) $D=-0.5$, (b) $D=0.1$, (c) $D=0.2$, (d) $D=0.5$. DMRG results are for $N=60$ sites  (straight lines); 
ED (dashed lines) for $N=15$ sites and PBC.}
\label{fig:1.5}
\end{figure}

\subsection{Magnetization process of frustrated $S$=1 chains with onsite anisotropy}

Most materials that realize $S$=1 chains typically also have a significant onsite anisotropy $D\not= 0$ of the form
$ D \sum_i (S_i^z)^2$ [see also
Eq.~(\ref{eq:2})]. 
We therefore study the effect of both frustration $J_2$ and a finite $D$ on the magnetization process of 
 frustrated $S$=1 chains. 
The problem of a spin-1 chain with frustration and onsite anisotropy $D\not= 0$ has previously attracted attention as well, 
however, for a slightly different model Hamiltonians than our Eq.~(\ref{eq:2}), including biquadratic terms of the form $(\mathbf S_i\cdot
\mathbf{S}_j)^2$.\cite{nakano98,kaburagi04}

Our DMRG and ED results for the magnetization process of  frustrated $S$=1 chains with a finite $D$ are presented in 
Fig.~\ref{fig:1.5} for fixed $J_2=0.6$ and $\Delta=1$. A negative $D$ does not affect the existence of the $M$=1/3 plateau 
state. We have verified the existence of the $1/3$ plateau  up to $D=-1$. 
While the width of the plateau is $\Delta h\approx0.80$ at $D=0$ and $J_2=0.6$, it increases steadily
with $D<0$. At $D=-0.5$, we obtain $\Delta h\approx 0.95$. Thus, a negative $D$ stabilizes the $M$=1/3 
plateau state. As an example, the magnetization curve $M(h)$ for $D=-0.5$ is shown in Fig.~\ref{fig:1.5}(a). 
A positive onsite anisotropy $D$, however, destroys the plateau state which disappears for $D\gtrsim 0.18$. At larger $D$, an anomaly around  $M$=2/3 emerges, as is shown in Fig.~\ref{fig:1.5}(d) 
for the case of $D=0.5$.
 Such a behavior at finite $D$ is expected since negative $D$ favors the states with a maximal projection $S_i^z$ on each site
(such as the {\it uud} state on the $M$=1/3 plateau). On the contrary, a positive $D$ plays the role of an easy-plane $XY$ 
anisotropy and thus suppresses the plateau formation.

\begin{figure}[t]
\centerline{\epsfig{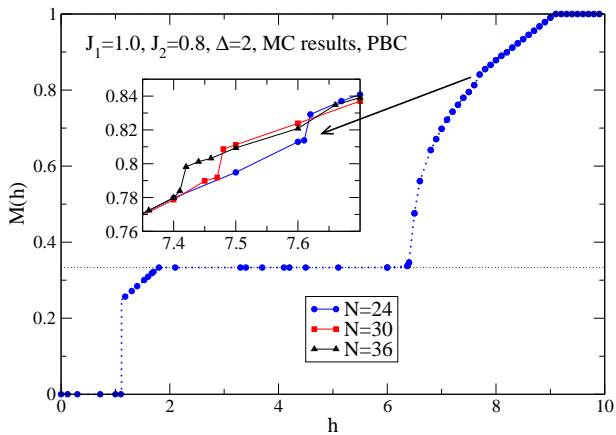}}
\caption{(Color online) 
Magnetization curve of a frustrated classical chain ($\Delta=2$) for $J_2=0.8$ obtained by means of MC simulations 
(circles). The horizontal, dotted line marks  $M$=1/3. 
The inset shows an enlarged view of the {\it kink} region around $h\approx 0.75$, 
illustrating the size  dependence.}
\label{fig:2.0}
\end{figure}

Let us summarize the main results of this Sec.~\ref{sec:2} on the magnetization process of frustrated chains with an isotropic
exchange. First, we have presented numerical results for $S$=1, 3/2, and 2 showing 
that a $M$=1/3 plateau is realized in the magnetization curve of  frustrated spin chains with isotropic exchange, while  a classical 
frustrated model does not support this state. Second, we have presented a magnetic phase diagram for frustrated $S$=1 and 3/2 chains with a 
special focus on the boundaries of the $M$=1/3 plateau region. A tendency is found that as $S>1/2$ grows, the width of the plateau
becomes more narrow. 

Moreover, kink  singularities exist in the magnetization curves of spin-1 chains  for $J_2/J_1\gtrsim 0.25$, which is -- 
at least close to the saturation field -- due to the emergence of two 
minima in the dispersion relation of the one-magnon excitations above the fully polarized state, similar to the case of 
$S$=1/2.\cite{okunishi99}  This effect further indicates the presence 
of an incommensurate region for fields $h>h_{\mathrm{kink}}$ above the high-field kink singularity. The inclusion of an onsite anisotropy term in the Hamiltonian of frustrated spin-$1$ chains
stabilizes the $M$=1/3-plateau  when the onsite anisotropy is negative ($D<0$), but the plateau is quickly destroyed by 
a positive $D\gtrsim 0.18$.

The magnetization curves of frustrated spin-3/2 chains exhibit a rich behavior with several kink singularities and the
$M$=1/3 plateau. Around the plateau and at small $J_2$, the magnetization increases in steps of $\Delta S=2S$. 
It is beyond the scope of this work to fully map out the phase diagram of spin-3/2 chains but interesting results are expected to emerge from future studies.


\section{The magnetization process of anisotropic frustrated spin chains in the easy-axis regime $\Delta>1$}
\label{sec:3}

In this section we compare our results for  frustrated spin-$S$ chains with $S>1/2$ to
a classical frustrated spin chain. Beyond a critical exchange anisotropy $\Delta_{\mathrm{c}}(J_2)$, the 
classical frustrated chain supports a $M$=1/3 plateau. 
Using linear spin-wave theory summarized in the Appendix, 
we derive several expressions for the phase boundaries
of the classical model  and in particular compare the phase boundaries of the plateau 
region to the cases of  $S$=1/2,1 and 3/2 chains. The quantum cases are  studied by means of ED and DMRG.
We find that, as $S$ increases, the classical phase boundaries are rapidly approached. 
Deviations between the quantum cases and the classical case are most significant close to the critical frustration for plateau formation.
The case of $S$=1/2  is clearly singled out: here, the critical frustration and the phase boundaries
are very different from spin $S$>1/2 chains and the classical model.
Finally, we briefly discuss additional phases of frustrated spin-1 chains in the easy-axis regime $\Delta>1$.

\begin{figure}[t]
\centerline{\epsfig{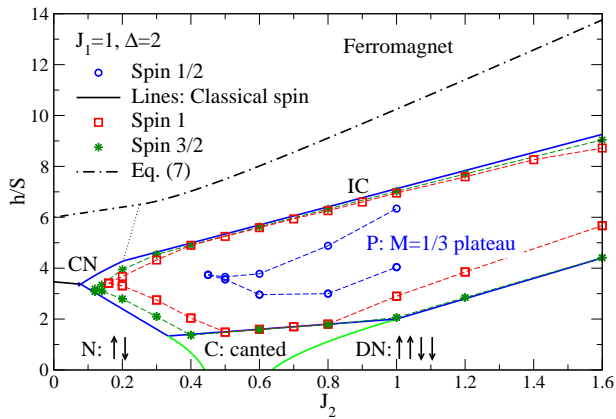}}
\caption{(Color online) 
 Magnetic phase diagram (field $h$ vs frustration $J_2$) 
 of a frustrated classical chain  at $\Delta$=2. 
 The phase boundaries of the classical model are given by  Eqs.~(\ref{eq:hab})--(\ref{eq:hdf}). They  are denoted by straight lines.
 The phases are labeled by capital letters (see Sec.~\ref{sec:class_phases}):
 {\bf N} is a N\'eel phase, {\bf C} is a canted phase with periodicity three, {\bf DN} is the double-N\'eel phase, 
 {\bf P} the $M$=1/3 plateau state, {\bf CN} is a canted N\'eel-like phase, and {\bf IC} is
 an incommensurate region.
 Numerical estimates of  the phase boundaries of the $M$=1/3 
 plateau region for spin 1/2,$1$, and 3/2 are denoted by circles, squares, stars, respectively. Dashed lines are guides to the eye.
 The dot-dashed line is the saturation field $h_{\mathrm{sat}}$ from Eq.~(\ref{eq:hsat}). 
 The dotted line schematically indicates the phase boundary between phases {\bf CN} and {\bf IC}, which we have not determined analytically.
 }
\label{fig:2.2}
\end{figure}

\subsection{Classical frustrated chain}
\label{sec:class}

We  perform MC simulations on chains of up to $N=36$ sites with periodic boundary conditions which allow us to 
determine the ground-state at finite magnetic fields as outlined in Sec.~\ref{sec:model}. Knowing the states gives us a handle on
analytical expressions for their energies.

The closely related problem of a classical Ising-like Heisenberg model on a triangular lattice was studied by 
Miyashita~\cite{miyashita86} and later on compared to 
$S$=1/2 Ising-like Heisenberg model on the  same lattice.\cite{nishimori86} Similar to the  Ising limit and spin 1/2,\cite{morita72}  the $M$=1/3 plateau was found and results for the magnetic 
phase diagram were reported. Here we focus on the one-dimensional case only, but we allow $J_1\not= J_2$ in our study, while 
Ref.~\onlinecite{miyashita86} concentrated on equal  exchange constants on all sides of the triangles. 

As an example for our MC calculations, we show the magnetization curve for $J_2=0.8$ and $\Delta=2$ in Fig.~\ref{fig:2.0}. At least five phases can be 
distinguished. First, a zero-field plateau exists, corresponding to a gap of $1.11$. The ground state
has an {\it up-up-down-down} ({\it uudd}) structure in the $S^z$ component, while at small $J_2$, a N\'eel state (\emph{up-down})
is realized. Larger $J_2$ emphasizes the double-chain character of our model. 
An {\it uudd} pattern results in N\'eel states on each of the single chains in the limit of $J_1\to 0$.

 \begin{figure}[t]
\centerline{\epsfig{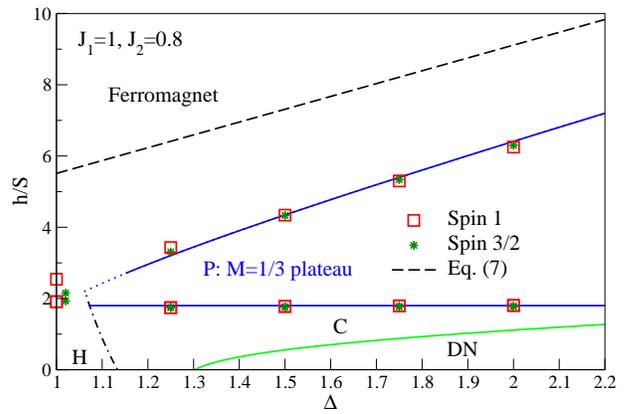}}
\caption{(Color online) 
 Magnetic phase diagram (field $h$ vs exchange anisotropy $\Delta$) 
 of a frustrated classical chain  at $J_2=0.8$ obtained by MC simulations. 
 The phase boundaries of the classical model are given by Eqs.~(\ref{eq:hab})--(\ref{eq:hdf}). They are  denoted by straight lines.
  Results for the phase boundaries of the $M$=1/3 
 plateau region for spin $1$ and spin $3/2$ from DMRG calculations are  included (squares: spin 1, stars: spin 3/2).
 The dashed line is the saturation field $h_{\mathrm{sat}}$ from Eq.~(\ref{eq:hsat}).
 Capital letters ({\bf C}, {\bf DN}, and {\bf P}) refer to phases shown in Fig.~\ref{fig:2.2} and are discussed in the text.
 {\bf H} denotes a helical phase.\cite{harada88}
 The dot-dashed line indicates the transition between phases {\bf H} and {\bf C}. Dotted lines are phase boundaries 
 that are only indicated schematically.
  }
\label{fig:2.3}
\end{figure}

Next, separated by a first-order transition the system enters a region where the magnetization increases linearly with $h$. 
This phase has periodicity $p=3$ and is a precursor of the plateau state: all spins lie in the same plane, one points exactly in 
the $-z$ direction while the other two are canted away from the $+z$ direction by  angles $\theta$ and $-\theta$ such that
\begin{equation}
\label{eq:theta}
\cos \theta = \frac 1{\Delta+1} \left(\frac h{J_1+J_2} +\Delta\right)\,. 
\end{equation}
As $h$ is increased, $\theta$ vanishes continuously, so that the transition into the plateau state is second order. 
The $M$=1/3 plateau exists for $J_1+J_2=1.8\leq h \lesssim 6.4108$. The system then enters a high-field region with a 
complicated dependence of $M$ on $h$. Within the error of our simulations the phase transition to this state from the plateau state is 
always second-order. The high-field state is probably incommensurate with the lattice period. This is illustrated by the
chain size dependence of the  magnetization curve and its small anomalies shown in the inset of Fig.~\ref{fig:2.0}.
Chains of finite length with PBC cannot accommodate a truly incommensurate state. This picture is  consistent with our 
analytical spin-wave calculations. From Eq.~(\ref{eq:1.2}) we obtain that, for the  present value of $J_2=0.8$ and 
$h = h_\text{sat} - 0^+$, the fully saturated state is instable against fluctuations with $\cos k = -J_1/4 J_2 = -0.3125$. On the
other hand, the plateau state is instable against the fluctuations with $\cos (3k) = 1$ as shown in the Appendix. This
dependence of $k$ on $h$ emphasizes
the incommensurability of the high-field phase. In this work, we have not attempted to study the $k(h)$ dependence or the nature of 
this state in further detail.

\subsection{The magnetic phase diagram of the classical model}

\label{sec:class_phases}

We  derive 
analytical expressions for the boundaries of the plateau of the classical model as well
as of several surrounding phases as a function of $\Delta$ and $J_2$. 
The results  are summarized in the magnetic phase diagrams in Figs.~\ref{fig:2.2} and \ref{fig:2.3}.

At low fields, we find three phases: a N\'eel state at small $J_2$ (region {\bf N} in Fig.~\ref{fig:2.2}), 
the canted commensurate phase  {\bf C}, and the double-N\'eel state {\bf DN} for $J_2\gtrsim 0.64$ with an {\it uudd} pattern.
According to Eq.~(\ref{eq:theta}), the magnetization of the \textbf{C} phase at $h=0$ is 
\begin{equation}
M(h=0)=\frac 1 3(2\cos\theta -1) = \frac{\Delta-1}{3 (\Delta +1)}\,, 
\end{equation}
which vanishes only in the isotropic limit $\Delta \to 1$. Therefore, as a  non-trivial consequence of the interplay
between frustration and exchange anisotropy, we observe 
a finite zero-field magnetization in a model with purely antiferromagnetic interactions.

The boundaries between the commensurate states mentioned above are given by:
\begin{widetext}
\begin{subequations}
\begin{eqnarray}
 h_{\mathrm{\bf N-C}}&=&\frac 1 2\left\{(J_1+J_2)(1-\Delta)+\sqrt{3}\sqrt{(J_1+J_2)(1+\Delta)\left[\Delta(3J_1-5J_2)-(J_1+J_2)\right]}\right\}\, ,
\label{eq:hab}\\
h_{\mathrm{\bf DN-C}} &=& \frac 1 2\left\{(J_1+J_2)(1-\Delta)+\sqrt{3}\sqrt{(J_1+J_2)(1+\Delta)\left[\Delta(3J_2-J_1)-(J_1+J_2)\right]}\right\} \,.\label{eq:hbc}
\end{eqnarray}
\end{subequations}
\end{widetext}

For the lower boundaries of the $M$=1/3 plateau we find:
\begin{subequations}
\begin{eqnarray}
h_{\mathrm{\bf N-P}} &=& 2\Delta(J_1-2 J_2);  \label{eq:had} \\
h_{\mathrm{\bf C-P}} &=& J_1+J_2;\label{eq:platlow} \\
 h_{\mathrm{\bf DN-P}} &=& \Delta (2J_2-J_1) \,.\label{eq:dnp}
\end{eqnarray}
\end{subequations}
Note that the critical field $h_{\mathrm{\bf C-P}}$ does not depend on $\Delta$.

Above the plateau, we have identified two phases.
At small $J_2$, a canted N\'eel state is realized (labeled {\bf CN} in Fig.~\ref{fig:2.2}), separated 
by a first-order transition line
\begin{equation}
 h_{\mathrm{\bf N-CN}} = 2 \sqrt{(J_1-J_2)(\Delta-1)[\Delta(J_1+J_2)+(J_1-J_2)]}  
\end{equation}
from the gapped N\'eel state {\bf N}. The \textbf{P} -- \textbf{CN} transition is also first-order. It occurs at
\begin{widetext}
\begin{equation}
h_{\mathrm{\bf P-CN}} = \frac{2}{3}  \lbrack J_1-J_2 + \Delta(J_1+J_2) + 2\sqrt{2 J_1 J_2 (2+\Delta^2)+J_1^2 (\Delta^2-\Delta -2) +J_2^2
(\Delta^2+\Delta -2)}\rbrack\, .
\label{eq:hde}
\end{equation}
\end{widetext}
For larger $J_2$ and magnetizations $M>1/3$, the situation becomes more complicated since incommensurability arises, as discussed
above. For the range of parameters that we study, the \textbf{P}-\textbf{IC} phase transition is always second order. Hence,
the phase boundary can be derived 
from an analysis of spin-wave instabilities  around the plateau state (see Appendix),
\begin{equation}
 h_{\mathrm{\bf P-IC}} =  \frac{J_1+J_2}2 [2\Delta - 1 + \sqrt{(4\Delta^2+4\Delta-7)}] \,.\label{eq:hdf}
\end{equation}
Equations (\ref{eq:had}) and (\ref{eq:hde}) result in an expression for the miminal frustration $J_{2,\mathrm{crit}}(\Delta)$ 
required for the formation of the $M$=1/3 plateau state in the vicinity of $\Delta=2$:
\begin{equation}
\frac{J_{\mathrm{2,crit}}}{J_1}= \frac{1-\Delta + 2\Delta^2-2\Delta \sqrt{\Delta(\Delta-1)}}{1-2\Delta+5 \Delta^2} \,.
\end{equation}
This implies $J_{2,\mathrm{crit}}\approx 0.079$ for the classical frustrated chain at $\Delta=2$.

To conclude this section, we present a cut through the phase diagram at fixed $J_2=0.8$ in Fig.~\ref{fig:2.3}, 
now plotting the magnetic field $h/S$ vs the exchange anisotropy $\Delta$. 
The double-N\'eel phase {\bf DN}  vanishes at $\Delta\approx 1.485$, and below this value of $\Delta$,
a helical phase {\bf H} emerges at zero magnetic field (see, e.g., Ref.~\onlinecite{harada88} for a discussion of helical order
in the isotropic case).
The $M$=1/3 plateau only exists above 
a critical value of the exchange anisotropy: $\Delta\approx 1.06$. We obtain this result by comparing
the energy of the helical phase {\bf H}, which -- according to our MC simulations -- extends up to the plateau at small $\Delta$,
to the energy of the plateau phase {\bf P}.
Note that a similar plot has been presented in Ref.~\onlinecite{miyashita86} for the triangular lattice with equal couplings $J$ 
on each triangle. A phase {\bf C} has not been reported in that work: at $J_1=J_2$, we consistently also
only find the double-N\'eel phase below the plateau, separated by a first-order transition from the plateau [see Fig.~\ref{fig:2.2}].

We emphasize that we have not yet fully explored all regions of the magnetic phase diagrams in Figs.~\ref{fig:2.2} and~\ref{fig:2.3}
of the classical frustrated chain. 
For instance, an analytical expression for the transition between phases {\bf CN} and {\bf IC} still needs to be found. Moreover,
additional phases may arise in the limit $J_2\gg J_1$. In this parameter region, we expect the plateau to vanish as the chains 
decouple.


\begin{figure}
\centerline{\epsfig{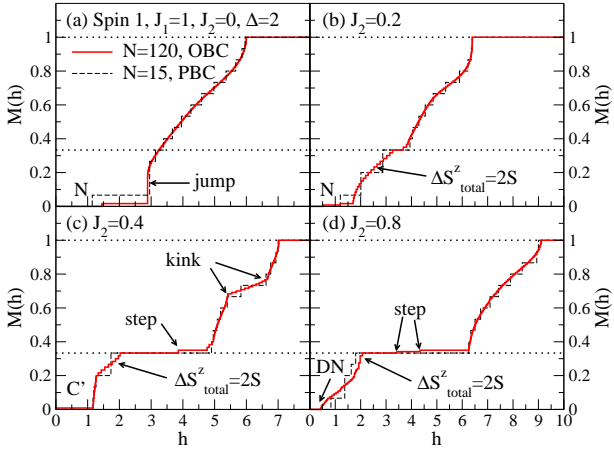}}
\caption{(Color online) 
 Magnetization curves  of frustrated spin-1 chains with an anisotropic exchange ($\Delta=2$) for (a) $J_2=0$,
 (b) $J_2=0.2$, (c) $J_2=0.4$, and (d) $J_2=0.8$. DMRG results (straight lines) are for $N=60$ sites, the dashed lines are ED results (PBC). The capital letters stand for: N\'eel phase {\bf N},
  canted phase {\bf C}, double-N\'eel phase {\bf DN}.  }
\label{fig:2.1}
\end{figure}

\subsection{$M$=1/3 plateau in the magnetization process of spin-1 and spin-3/2 chains: Comparison to the classical case}

\begin{figure}
\centerline{\epsfig{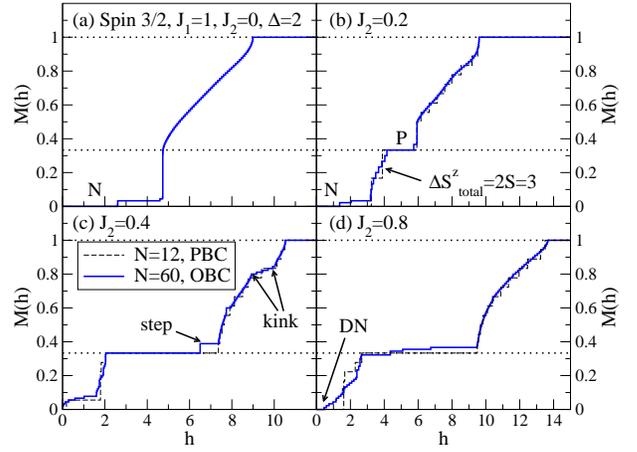}}
\caption{(Color online) 
 Magnetization curves for  frustrated spin-3/2 chains with an anisotropic exchange ($\Delta=2$) for (a) $J_2=0$,
 (b) $J_2=0.2$, (c) $J_2=0.4$, and (d) $J_2=0.8$. DMRG results (straight lines) are for $N=60$ sites, 
 the dashed lines are ED results (PBC). The capital letters stand for: N\'eel phase {\bf N},
  canted phase {\bf C}, double-N\'eel phase {\bf DN}.}
\label{fig:2.5}
\end{figure}

We now turn to the magnetization process of frustrated  quantum spin chains with $S$=1 and $3/2$ in the easy-axis
regime $\Delta>1$. Numerical results for  $M(h)$ are presented in Figs.~\ref{fig:2.1} and~\ref{fig:2.5} 
for $J_2=0,0.2,0.4,0.8 $ and $\Delta=2$.
We briefly discuss the magnetization process and emergent phases starting with the low-field part. 
The focus here is on to what extent the $S$=1 and 3/2 chains resemble the phases found for the classical
chain. Results on the  zero-field phase diagram of frustrated spin-1 chains and $\Delta>1$ 
can be found  in Refs.~\onlinecite{hikihara00} and \onlinecite{murashima05}. At a finite $\Delta>1$ and for spin-1,
essentially five phases have been identified: a N\'eel state at  small $J_2$ and $\Delta> 1.18$, the Haldane phase, the Double Haldane
phase and the Double-N\'eel phase for $\Delta \gtrsim 1.95$ on the large $J_2$ side (see Ref.~\onlinecite{hikihara00} and further references
therein). Our results reported below are in agreement with this picture.

We now concentrate on the example of $\Delta$=2. First, at small $J_2$, both the $S$=1 and 3/2 chains are in the N\'eel phase {\bf N}, see for instance 
Figs.~\ref{fig:2.1}(a) and \ref{fig:2.5}(a). 
This phase is gapped and therefore, a zero-field plateau exists 
in the magnetization curves. The N\'eel phase terminates at $J_{2,\mathrm{\bf N-C'}}\approx 0.29$ for $S$=1 and 
$J_{2,\mathrm{\bf N-C'}}\approx 0.39$ for $S$=3/2. 
For larger $J_2$, an intermediate region {\bf C'} follows, which is gapped in the case of 
$S$=1 [see Fig.~\ref{fig:2.1}(c)]. 
This region must contain both the Haldane and the Double-Haldane phase, with a transition at
$J_2\approx 0.8$.\cite{hikihara00}
The commensurabilty and the existence of a gap in the case of $S$=3/2 in this region still needs to 
be clarified. The width in $J_2$ of this region at zero field becomes narrower as $S$ grows, approaching the boundaries of the
phase {\bf C} of the classical model (see Fig.~\ref{fig:2.2}).
At $J_{2,\mathrm{\bf C'-DN}}\approx 1.02$ for $S$=1 and $J_{2,\mathrm{\bf C'-DN}}\approx 0.94$ for $S$=3/2, the systems undergoes  another first-order
transition and enters the double-N\'eel phase {\bf DN},
which is gapped for both values of $S$ (see Figs.~\ref{fig:2.1}(d) and \ref{fig:2.5}(d), $J_2=0.8$).

At small $J_2$ and $h>0$, the \textbf{N} phase terminates at a first-order transition, indicated by  macroscopic jumps
in the magnetization curves of both $S$=1 and 3/2 chains  [see the case of $J_2=0$ in Figs.~\ref{fig:2.1}(a) and \ref{fig:2.5}(a)].
The critical frustrations for the formation of the $M$=1/3 plateau state are $J_{2,\mathrm{crit}}\approx 0.16$ for $S$=1
and $J_{2,\mathrm{crit}}\approx 0.12$ for $S$=3/2. Interestingly, 
an EO region 
($\Delta S^z_{\mathrm{total}}=2$) exists around the plateau at small $J_2$ in the case of $S$=1 [see Fig.~\ref{fig:2.1}(b) and (c) 
for $J_2=0.2$ and $J_2=0.4$]
and extends far below the plateau. 
For $S$=3/2, we find a corresponding region with $\Delta S^z_{\mathrm{total}}=3 $. 
An example is shown in Fig.~\ref{fig:2.5}~(b) for $J_2=0.2$. 
We are thus led  to conjecture that
frustrated spin-$S$ chains generally realize regions with $\Delta S^z_{\mathrm{total}}=2S$. We have verified that this is the case for 
spin $S$=2 as well. 

The $M$=1/3 plateau becomes very broad with increasing $J_2$, as is evident from the case of $J_2=0.4$ depicted in Figs.~\ref{fig:2.1}(c) 
and \ref{fig:2.5}(c). Similar to the case of isotropic chains, boundary induced steps appear on the plateau. We have not
determined the upper critical frustrations for the plateau formation, but we did verify that 
the plateau disappears for large $J_2$ in both the cases of $S$=1 and 3/2.
For magnetizations $M>1/3$, we mention that kink singularities are again found for $J_2 > 0.25$, see the examples in Figs.~\ref{fig:2.1}(c) and \ref{fig:2.5}(c).

We proceed  with a comparison of the critical frustrations $J_2$ of the $M$=1/3 plateau state
as a function of $S$, listed in  Table~\ref{tab:2}. For $S>1/2$, a clear trend towards the classical result is seen.
Numerical results for the phase boundaries of the plateau state for both $S$=1 (squares) and $S$=3/2 (stars) are included in 
Figs.~\ref{fig:2.2} ($h$ vs  $J_2$ at $\Delta=2$) and \ref{fig:2.3} ($h$ vs $\Delta$ at $J_2=0.8$).
It turns out that the phase boundaries in Fig.~\ref{fig:2.2} converge rapidly with increasing $S$ towards the classical result.
For instance, on the scale of Fig.~\ref{fig:2.2}, the classical result for the transition $h_{\bf C-P}$ from the phase  {\bf C} 
to the plateau cannot be distinguished from the quantum cases. The data points for $S$=1 on the boundary $h_{\bf C-P}$ quantitatively 
deviate from the classical result, but the data points for $S$=3/2 are already very close to the classical result
for $J_2\approx 0.6$. Deviations are strongest in both the limit of small and large $J_2$. 

With respect to Fig.~\ref{fig:2.3} ($h$ vs $\Delta$ at $J_2=0.8$), we note that 
sizable deviations between the plateau boundaries of the classical chain and
the quantum cases only appear at small exchange anisotropies close to $\Delta=1$.
While the spin-1 chain realizes the plateau in the isotropic limit $\Delta=1$,
the plateau only opens in $M(h)$ of spin-3/2 chains beyond a critical $1 < \Delta_{\mathrm{crit,3/2}} \lesssim 1.02 <
\Delta_{\mathrm{crit,class}}$. 

We can finally conclude that  the magnetic phase diagram of frustrated $S$=1 and 3/2 chains
in the easy-axis regime strongly resembles the phases found in the classical limit. Examples are the N\'eel phase {\bf N},
the double-N\'eel phase {\bf DN}, and most importantly, the $M$=1/3 plateau state. In the latter case,
the similarity is not only of qualitative nature, but the phase boundaries of the plateau rapidly approach the classical result 
with increasing $S$. We emphasize our findings on increases of the magnetization in steps corresponding to 
$\Delta S_{\mathrm{total}}= 2S$ which may indicate interesting binding effects between excitations.

\begin{table}
\begin{ruledtabular}
\begin{tabular}{cc}
Spin $S$  & $J_{2,\mathrm{crit}}/J_1$  \\\hline
1/2 &  0.45 \\
1   &  0.16 \\
3/2 &  0.12 \\
2   &  0.10 \\
$\infty$ &  0.079
\end{tabular}
\end{ruledtabular}
\caption{Easy-axis regime, $\Delta=2$: Lower critical $J_{2,\mathrm{crit}}$ for the formation
of the $M$=1/3 plateau for spin 1/2, 
 1, 3/2, and 2 (DMRG results) and the classical limit $S\to\infty$.  
}\label{tab:2}
\end{table}



\section{Summary}
In this work we have studied the magnetization process of 
frustrated spin chains. Our numerical results using DMRG and ED techniques reveal a complex and rich 
behavior of the magnetization process of $S$=1, 3/2, and 2 chains. We find jumps, zero- and finite-field 
plateau states, as well as kink singularities.  Our primary result is the  $M$=1/3 plateau  state with 
broken translational invariance and an {\it up-up-down} pattern in the spin component parallel
to the field. 
 The 
existence of this plateau indicates the presence of a gap in the spectrum of excitations at finite magnetic fields.
We have numerically estimated the phase boundaries of the plateau for $S$=1 and 3/2 chains with isotropic exchange interactions. 
An additional onsite anisotropy term $D (S_i^z)^2$, necessary for the description of many Haldane materials,
stabilizes the plateau, when $D$ is negative but rapidly destroys the plateau state when $D$ is positive.

The $M$=1/3 plateau state can be followed down from the Ising-limit by decreasing the ratio $\Delta/J_{1}$ for fixed $S$.
For that reason and for its N\'eel-type of order, it has been classified as a classical 
state,\cite{okunishi03,hida05}  but quantum fluctuations seem to stabilize it 
in the case of isotropic exchange: a classical frustrated chain does not show a plateau here.
We have studied the classical frustrated spin chain in detail, which realizes the $M$=1/3 plateau in the easy-axis regime. 
Our work provides several analytical results
for the magnetic phase diagram of the classical frustrated chain including the phase boundaries of the $M$=1/3 plateau for $\Delta>1$.
In the easy-axis regime, the phase diagrams of frustrated $S$=1 and 3/2 chains do  not only qualitatively
resemble the classical result, but also the phase boundaries of the $M$=1/3 plateau rapidly approach the classical result with 
increasing $S$. The extreme quantum case of $S$=1/2 is singled out, 
the critical frustrations are  different from  $S>1/2$ and the classical cases both in the isotropic ($\Delta=1$) and the easy-axis regime.
Our conjecture is that the plateau exists for all frustrated spin-$S$ chains in the easy-axis regime, 
as our results for  $S$=1, 3/2,  2 and the classical case suggest.

Our results on the emergence of the $M$=1/3 plateau may be of relevance to the materials mentioned in the introduction that
are suggested to realize zigzag spin-1 chains:  CaV$_2$O$_5$ (Ref.~\onlinecite{kikuchi01}),
NaR(WO$_4$)$_2$ (Ref.~\onlinecite{masuda02}), and  Tl$_2$Ru$_2$O$_7$ (Ref.~\onlinecite{lee06}).
Moreover, the $M$=1/3 plateau can be expected to emerge in the magnetization process of all materials that realize a spin-$S$ Heisenberg model
on a triangular lattice, as is well known for spin 1/2 (see, e.g., Ref.~\onlinecite{chubokov91}). Indeed, plateaux at both $M$=1/3 and $M$=2/3 have recently been reported for the $S$=1 bilinear-biquadratic
Heisenberg model on the triangular lattice.\cite{laeuchli06} 

As by-products to the plateau study, and in particular close to the plateau region, we have identified
additional phases. We have obtained  the phase diagram 
of the frustrated spin-1 chain with isotropic interactions, extending the results of
other studies.\cite{kolezhuk05}
We highlight another interesting finding: all frustrated spin-$S$ chains seem to realize regions where the 
magnetization increases in steps  corresponding to 
$\Delta S_{\mathrm{total}}= 2S$.
Extensions of our work will comprise a full characterization of different phases in terms of correlation functions and
a more detailed study of excitations. A timely subject is the emergence of chirally ordered states in frustrated quantum magnets, which 
is investigated both experimentally\cite{kikuchi01,affronte99} and
theoretically.\cite{kolezhuk05,kaburagi99,hikihara00,hikihara01} We hope that our work will 
stimulate future research activities in these directions.

\indent {\bf Acknowledgments - } It is a pleasure to thank D.C. Cabra, A. Honecker, U. Schollw\"ock, and T. Vekua for fruitful discussions.
We further thank A. Kolezhuk and U. Schollw\"ock for sending us their numerical data for the spin gap of the frustrated spin-1 chain for comparison.
Research at ORNL is sponsored by the Division of Materials Sciences and Engineering,
Office of Basic Energy Sciences, U.S. Department of Energy, under contract DE-AC05-00OR22725 with Oak
Ridge National Laboratory, managed and operated by UT-Battelle, LLC.
I.S. is supported in part by NSF grant DMR-0072998.
E.D. and F.H.-M.  are supported in part by  NSF grant DMR-0443144.

\begin{figure}[t]
\centerline{\epsfig{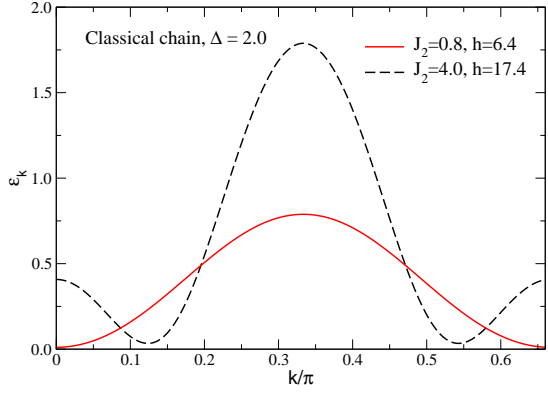}}
\caption{\label{fig:app}Lowest branch of spin-wave excitations in the \emph{uud} plateau state close to instability.
The solid line is for the range of parameters studied in Sec.~\ref{sec:3}, showing instability towards $k=0$ fluctuations. 
The dashed line is for a larger $J_2$ with an incommensurate instability.
}
\end{figure}
\appendix

\section{Spin waves in the $M$=1/3 plateau state}

In this appendix, we outline the linear spin-wave theory that allows us to determine the phase boundaries of
the classical model. In this appendix, $D=0$.
We start with the equation of motion for the spin operators $\mathbf S_p$ on the  $p$th site of the chain, 
$d \mathbf S_p /d t = i [H, \mathbf S_p]$,
where $H$ is the Hamiltonian Eq.~(\ref{eq:2}) and the brackets denote the commutator. Explicitly, we obtain the following,
\begin{eqnarray}
\hspace{-0.5cm}\frac {d S_p^x}{d t} &=& S_p^z \sum_q J_{pq} S_q^y - \Delta\, S_p^y \sum_q J_{pq} S_q^z + h S_p^y\nonumber\\
\hspace{-0.5cm}\frac {d S_p^y}{d t} &=& -S_p^z \sum_q J_{pq} S_q^x + \Delta\, S_p^x \sum_q J_{pq} S_q^z - h S_p^x,
\label{eq:app}
\end{eqnarray}
where $J_{p,q} = J_1$ for $q=p\pm 1$, $J_2$ for $q=p\pm 2$, and zero otherwise. The fluctuations in $S^z$ are negligible since 
$[H, \mathbf S_p^z]$ contains only products of the form $S_p^xS_q^y$ which are of second-order in small fluctuations.
Here, we are only interested in the classical limit which greatly simplifies the  calculations. Hence, in the following, 
$\mathbf S_p$ are viewed as three-dimensional classical vectors of length $S$.

It is convenient to divide the \emph{uud} plateau structure into ``unit cells'', each containing three chain sites. To this end,
we replace $p$ by a combined index $(m,\alpha)$, where $p=3 m + \alpha$, $m = 0,...,N/3-1$ is a unit cell number, and $\alpha= 0, 1, 2$ 
enumerates the sites within the unit cell. Correspondingly, in the linear approximation, $S^z_{m,0}=S^z_{m,1}\approx S$ and 
$S^z_{m,2}\approx -S$, and~(\ref{eq:app}) becomes a
system of linear differential equations for $S^x_{m,\alpha}$ and $S^y_{m,\alpha}$. Introducing the plane-wave solution 
$S^{x,y}_{m,\alpha} = S^{x,y}_{\alpha} \exp [i (k\, p - \epsilon\, t)]$ and a vector notation for the amplitudes
${\cal S} = (S^x_0, S^y_0, S^x_1, S^y_1, S^x_2, S^y_2)$, Eq.~(\ref{eq:app}) can be written as $A \cdot {\cal S}=0$,
where ($\gamma \equiv e^{i k}$)
\begin{widetext}
\begin{equation}
A = 
\begin{pmatrix}
i \epsilon & h & 0 & J_1 \gamma + J_2 \gamma^{*2} & 0 & J_1 \gamma^* + J_2 \gamma^2 \\
-h & i \epsilon & - J_1 \gamma - J_2 \gamma^{*2} & 0 & - J_1 \gamma^* - J_2 \gamma^2 & 0 \\
0 & J_1 \gamma^* + J_2 \gamma^2 & i \epsilon & h & 0 & J_1 \gamma + J_2 \gamma^{*2} \\
-J_1 \gamma^* - J_2 \gamma^2 & 0 & -h & i \epsilon & -J_1 \gamma - J_2 \gamma^{*2} & 0\\
0 & - J_1 \gamma^* - J_2 \gamma^{*2} & 0 & -J_1 \gamma^* - J_2 \gamma^2 & i \epsilon & h - 2\Delta(J_1+J_2)\\
J_1 \gamma + J_2 \gamma^{*2} & 0 & J_1 \gamma^* + J_2 \gamma^2 & 0 & -h + 2\Delta(J_1+J_2) & i \epsilon
\end{pmatrix} \,.
\label{eq:det}
\end{equation}
\end{widetext}
The spin-wave spectrum $\epsilon_k$ is defined as the solution to the equation 
\begin{equation}
f(\epsilon_k, k) \equiv \det A=0. 
\label{eq:fdef}
\end{equation}
We do not give the explicit cumbersome expression for $f(\epsilon, k)$. It turns out, that, as a function of $k$, $f(\epsilon, k)$ only 
depends on $\cos (3k)$ which is consistent with the broken translational symmetry of the plateau state. 
In general, $f(\epsilon, k)$ is a third-order polynomial in $\epsilon^2$, 
the roots of which cannot be written in a closed analytical form. We can, however, investigate the minima of the function 
$\epsilon_k$ that is implicitly  given by the equation $f(\epsilon, k) = 0$. 
The extrema of $\epsilon_k$ are defined by
\begin{equation}
\frac{d\epsilon}{d k}= 
-\frac {\partial f/\partial k}{\partial f/\partial \epsilon} = 0\,.
\end{equation}
The plateau state becomes instable towards  spin-wave fluctuations when $\epsilon^\text{min}_k \to 0$.
After some algebra, we obtain that the corresponding wave vectors are defined by $\cos (3k) = 1 $
or
\begin{equation}
\cos (3k) = \frac{J_1}{4J_2^3} [h J_2 + 2\Delta J_2(J_1+J_2)-J_1^2-3J_2^2)]\,. \label{eq:crit_k_inc}
\end{equation}
For the range of parameters addressed in Sec.~\ref{sec:3}, the incommensurate solution Eq.~(\ref{eq:crit_k_inc}) is not realized, and the critical fluctuations
correspond to the first solution.
 As an example,  the lowest spin-wave dispersion branch, corresponding to the parameters close to 
the instability, is shown in Fig.~\ref{fig:app}. However, for higher $J_2$, the incommensurate fluctuations may become critical 
as is shown by the dashed line in Fig.~\ref{fig:app}. A more thorough study of that region is beyond the scope of the present paper. 

Substituting~ $\cos (3k) = 1$ and $\epsilon_k = 0$ in Eq.~(\ref{eq:fdef}), we obtain the  boundaries of the plateau state, by
checking the stability against small fluctuations. The solutions of Eq.~(\ref{solu}) [see below]
yield the second-order transition boundaries~(\ref{eq:platlow}) and~(\ref{eq:hdf}).
\begin{widetext}
\begin{equation}
(h-J_1-J_2)^2 
[h^2 + h (J_1+J_2)(1-2\Delta) + 2 (1-\Delta) (J_1+J_2)^2]^2 = 0\,.\label{solu}
\end{equation}
\end{widetext}


\bibliographystyle{/home/9fh/bibliothek/pf}
\bibliography{/home/9fh/bibliothek/bbase}

\begin{thebibliography}{10}
\bibitem{endnote63}
Note that the (staggered) local moment $|S_i^z |/S$ 
increases with increasing spin-$S$, indicating 
the expected suppression of quantum fluctuations 
as $S$ approaches the classical limit. In the case of 
spin 2 shown in Fig.\protect \nobreakspace  {}\ref {fig:1.1}
\protect \nobreakspace  {}(c), the polarization $| S_i^z |/S$ 
has not yet reached its maximum value, which -- 
deep in the plateau region -- is larger than the corresponding 
local moment realized for spin 1 and 3/2.
\end{thebibliography}


\end{document}